\documentstyle[12pt,a4]{article}

\begin{document}

%%%%% title %%%%%%%%%%%%%%%%%%%%%%%%%%%%%%%%%%%%%%%

\title{Mass Matrix of Majorana Neutrinos}
\author{Takeshi FUKUYAMA \thanks{e-mail: fukuyama@bkc.ritsumei.ac.jp} and 
Hiroyuki NISHIURA \thanks{e-mail: nishiura@jc.oit.ac.jp}  
                   \\
        $\ast$~~Department of Physics, \\
        Ritsumeikan University, Kusatsu, \\
        Shiga, 525 Japan \\
        $\dagger$~~ Department of General Education, \\
        Junior College of Osaka Institute of Technology, \\
        Asahi-ku, Osaka,535 Japan}
\date{}

\maketitle
 
%%%%%%%%%%%%%%% abstract %%%%%%%%%%%%%%%%%%%%%%%%%%%%%%%%%%%%

\begin{abstract}
We present a massive Majorana neutrino model and see how it is constrained from 
the solar and atmospheric neutrino deficit experiments.  This model incorporates 
the seesaw mechanism and Peccei-Quinn symmetry.  Its consequence to the 
neutrinoless double beta decay is also discussed.

\vskip 1cm
{\bf Key words :} Majorana particle,
                  Seesaw mechanism, Peccei-Quinn symmetry 

\end{abstract}

%\vskip 2cm
PACS number  12.15.Ef,14.60.Gh
\vskip 7cm
%\footnote{e-mail: fukuyama@bkc.ritsumei.ac.jp}
\vfill\eject

%%%%%%%%%%%%%%%%%%%%%%%%%%%%%%%%%%%%%%%%%%%%%%%%%%

%%%%%%%%%%%%%%%%%%%%%%%%%%%%%%%%%%%%%%%%%%%%%%%%%%%%%%%%%%%%%%%%%%%%%%%%%%
%\section{Introduction}
From the recent solar neutrino and atmospheric neutrino experiments 
\cite{Kamioka} \cite{Homestake} \cite{Gallex} \cite{Sage}, it becomes very 
probable that the neutrinos have masses. 
In this letter, we propose the model of massive Majorana neutrino and its 
physical consequences,especially to neutrinoless double beta decays. 
\par
Our physical standpoints are as follows.  The solar (atmospheric) neutrino 
deficit is due to $\nu _e -\nu _{\mu}(\nu _{\mu}-\nu _{\tau})$ oscillation and 
all neutrinos are of Majorana type.  Their masses are generated by the seesaw 
mechanism \cite{yana} \cite{gell} and have the hierarchy of
\begin{equation}
m_{\nu _e}={m_e ^2 \over M_R},~~m_{\nu _{\mu}}={m_{\mu}^2 \over M_R},~~ m_{\nu 
_{\tau}}={{m_\tau}^2\over M_R}
\end{equation}
with $M_R$ is the order of the Peccei-Quinn symmetry breaking.  Our strategy is, 
therefore, to construct a model which realizes the above mentioned standpoints. 
  
\par
Mass Lagrangian for the leptonic part includes two SU(2) doublets 
$\phi_1$,$\phi_2$ and one singlet $\phi_3$ Higgs fields.  Its explicit form is 
given by 
\begin{equation}
-L_{mass}=\sum_{i,j}^{3}  f_{ij}^{(l)} \bar l_L^{(i)} \phi_1 
e_R^{(j)}+\sum_{i,j}^{3}f_{ij}^{(\nu )} \bar 
l_L^{(i)}\phi_2\nu_R^{(j)}+\sum_{i,j}^{3}f_{ij}^{(M)}\bar \nu_R^{c (i)} 
\nu_R^{(j)} \phi_3+h.c.
\label{model}
\end{equation}
Here $(i,j)$ is the generation. $l_L$ is the left-handed doublet and 
$e_R^{(i)}(\nu_R^{(i)})$ is righ-thanded charged lepton (neutrino) singlet of  
i'th generation.  The third term represents massive Majorana neutrinos which 
induce the seesaw mechanism. This Lagrangian has the local SU(2) $\times$ U(1) 
symmetry and global Peccei-Quinn and lepton number symmetries.
\par
At this stage we have no relation among the coupling constants though Eq.(1) 
suggests some relations.  Later we will see how the present experiments 
constrain these parameters.  Here we study first the structure of mass 
Lagrangian after the spontaneous symmetry breaking 
\cite{Dine}\cite{Kim}\cite{Fukugita}. We expand Higgs fields as
\begin{eqnarray}
\phi_1&=&{1\over \sqrt{2}} \left({\xi_1^y+i\xi_1^z \over 2} \atop 
{\rho_1+\chi_1+{i \xi_1^x \over 2}}\right),\nonumber \\
\phi_2&=&{1\over \sqrt{2}} \left({\rho_2+\chi_2+{i\xi_2^x \over 2}} \atop 
{\xi_2^y+i\xi_2^z\over 2}\right), \\
\phi_3&=&{1\over \sqrt{2}} \left(\rho_3+\chi_3+{i\xi_3\over 2}\right), \nonumber
\end{eqnarray}
where $\rho_i$ are the vacuum expectation values.  Then the following 
combinations of $\xi_i^j$ are gauged away by the weak boson transformations,
\begin{eqnarray}
Z_\mu'&=&Z_\mu-{1\over \sqrt{g^2+g'^2}\rho}\partial_\mu\phi^Z,\nonumber\\
W_\mu'&=&W_\mu-{1\over \rho g}\partial_\mu ({\phi^{W_2}+i\phi^{W_1}\over 
\sqrt{2}}),\\
\bar {W}_\mu'&=&\bar {W}_\mu-{1\over \rho g}\partial_\mu 
({\phi^{W_2}-i\phi^{W_1}\over \sqrt{2}})\nonumber.
\end{eqnarray}

Here
\begin{eqnarray}
\phi^Z&\equiv&{\rho_2\xi_2^x-\rho_1\xi_1^x \over \rho},\nonumber \\
\phi^{W_1}&\equiv&{\rho_2\xi_2^y-\rho_1\xi_1^y\over \rho}, \\
\phi^{W_2}&\equiv&{\rho_2\xi_2^z+\rho_1\xi_1^z\over \rho}\nonumber
\end{eqnarray}
with $\rho\equiv \sqrt{\rho_1^2+\rho_2^2}$. Here we have written the 
infinitesimal trnsformations, sufficient to see the Higgs mechanism.
\par
Thus $\varphi^Z,\varphi^{W_1}$ and $\varphi^{W_2}$ defined below remain as the 
dynamical variables together with $\chi_i$,
\begin{eqnarray}
\varphi^Z&\equiv &{\rho_2\xi_1^x-\rho_1\xi_2^x\over \rho},\nonumber\\
\varphi^{W_1}&\equiv &{\rho_2\xi_1^y+\rho_1\xi_2^y\over \rho},\\
\varphi^{W_2}&\equiv &{\rho_2\xi_1^z-\rho_1\xi_2^z\over \rho}.\nonumber
\end{eqnarray}
So far we have not been able to restrict the parameters in Eq.(\ref{model}).  
Fritzsch assumed some additional symmetries and predicted  the flavour mixing 
angles in quark sector \cite{Fritzsch}.  Our procedures reverse this process in 
lepton sector.  That is, we consider first how the experiments constrain, 
especially, the neutrino mass matrix  $M_{light}$ after the seesaw mechanism,
\begin{equation}
M_{light}=-M_D M_R^{-1}M_D ^T ,
\label{light M}
\end{equation}  
where$(M_D)_{ij}\equiv f_{ij}^{(\nu )}\rho _2$ and $(M_R )_{ij}\equiv 
f_{ij}^{(M)}\rho _3$.
\par
If we adopt that the atmospheric neutrino deficit is due to $\nu_{\mu} 
-\nu_{\tau}$ oscillation, $\theta_2 (\theta_2\equiv \theta_{23}$ and analogously 
$\theta_3\equiv \theta_{31},\theta_1\equiv \theta_{12}$) may be constrained to 
be $\theta_2\sim {\pi\over 4}$ from the experiment \cite{Fukuda}.
Also we assume that the solar neutrino deficit is due to $\nu_e -\nu_{\mu}$ 
oscillation and set $\theta _3 \sim 0$.  That is, the orthogonal (we have not 
considered CP violation phases )  lepton mixing  matrix
 defined by
\begin{equation}
U=
\left(
\begin{array}{ccc}
1&0&0\\
0&c_2&s_2\\
0&-s_2&c_2
\end{array}
\right)
\left(
\begin{array}{ccc}
c_3&0&s_3\\
0&1&0\\
-s_3&0&c_3
\end{array}
\right)
\left(
\begin{array}{ccc}
c_1&s_1&0\\
-s_1&c_1&0\\
0&0&1
\end{array}
\right)
\end{equation}
becomes
\begin{equation}
U=\left(
\begin{array}{ccc}
c_1&s_1&0\\
-{1\over \sqrt{2}}s_1&{1\over \sqrt{2}}c_1&{1\over\sqrt{2}}\\
{1\over\sqrt{2}}s_1&-{1\over\sqrt{2}}c_1&{1\over\sqrt{2}}
\end{array}
\right),
\label{KM}
\end{equation}
where $s_i$ ($c_i$)is $sin \theta_i$ ($cos \theta_i$).  $M_{light}$ is 
constrained to be
\begin{eqnarray}
M_{light}&=&U
\left(
\begin{array}{ccc}
-m_1&0&0\\
0&m_2&0\\
0&0&m_3
\end{array}
\right)
U^T\nonumber\\
&=&\left(
\begin{array}{ccc}
-c_1^2m_1+s_1^2m_2&{1\over\sqrt{2}}c_1s_1(m_1+m_2)&-{1\over\sqrt{2}}c_1s_1(m_1+m
_2)\\
{1\over\sqrt{2}}c_1s_1(m_1+m_2)&{1\over 2}(-s_1^2m_1+c_1^2m_2+m_3)&{1\over 
2}(s_1^2m_1-c_1^2m_2+m_3)\\
-{1\over\sqrt{2}}c_1s_1(m_1+m_2)&{1\over 2}(s_1^2m_1-c_1^2m_2+m_3)&{1\over 
2}(-s_1^2m_1+c_1^2m_2+m_3)
\end{array}
\right).\nonumber\\
\end{eqnarray}
Here we have set the sign of $m_1$ negative. Indeed we can always change the 
sign of the mass by making the transformation $\psi_R\rightarrow -\psi_R$ and 
$\psi_L\rightarrow \psi _L$. There exists a mass hierarchy of $m_1\ll m_2\ll 
m_3$ and we have no lower bound with respect to $m_1$ from the neutrino 
anomalies.  So we assume
\begin{equation}
-c_1^2m_1+s_1^2m_2=0
\end{equation}
That is, the lightest neutrino mass is considered to be generated only by the 
flavour mixing.  Eq.(11) shows that we adopt the small angle solution for the 
solar neutrino oscillation. Then $M_{light}$ is reduced to
\begin{equation}
M_{light}=
\left(
\begin{array}{ccc}
0&A&-A\\
A&B&C\\
-A&C&B
\end{array}
\right),
\label{form}
\end{equation}
where
$$A\equiv {1\over \sqrt{2}}\sqrt{m_1m_2},~~B\equiv {1\over 2}(-m_1+m_2+m_3)$$ 
and $$C\equiv {1\over 2}(m_1-m_2+m_3).$$  It may be interesting to compare this 
with the quark mass matrix proposed by Fritzsch \cite {Fritzsch}.  Then we will 
see how $M_{light}$ in Eq.(\ref{form}) is affected on seesaw mechanics (7).
\par
Firstly, let us assume that $M_D$ in Eq.(7) is diagonalized as $M_D\propto 
diag(m_e, m_{\mu} , m_{\tau} )$ for simplicity. That is,$$f_{ij}^{(\nu )}\phi 
_2=\alpha ~diag(m_e,m_{\mu} ,m_{\tau} )$$ in Eq.(2), where $\alpha$ is some 
constant.  You should be careful not to confuse this with $f_{ij}^{(l)}\phi _1$. 
 The reason for this choice is to realize Eq.(1) naively. In this case 
$f_{i,j}^{(l)}\phi_1=diag(m_e, m_{\mu} , m_{\tau})$ also must be satisfied.
Then from Eqs.(7) and (9) we obtain the matrix $M_R$ as 
\begin{equation}
M_{R}=-\alpha^2
\left(
\begin{array}{ccc}
m_e^2{m_1-m_2 \over m_1 m_2}&m_em_{\mu}\sqrt{{1\over 
2m_1m_2}}&-m_em_\tau\sqrt{{1\over 2m_1m_2}}\\
m_em_{\mu}\sqrt{{1\over 2m_1m_2}}&{m_\mu^2\over 2m_3}&{m_\mu m_\tau \over 
2m_3}\\
-m_e m_\tau \sqrt{{1\over 2m_1 m_2}}&{m_\mu m_\tau \over 2m_3}&{m_{\tau}^2 \over 
2m_3}
\end{array}
\right)
\end{equation}
This has a rather complicated structure and is unlikely to posess some symmetry. 
 So we adopt the other option that $M_D$ and $M_R$ have the same structure as 
$M_{light}$ in Eq.(\ref {form}).  Same structure means the same relationships 
between the components of the respective matrix. It is remarkable that this 
assumption is consistent with seesaw mechanism (7).  That is, if we accept
\begin{equation}
M_D=\left(
\begin{array}{ccc}
0&A_D&-A_D\\
A_D&B_D&C_D\\
-A_D&C_D&B_D
\end{array}
\right),\\~~
M_R=\left(
\begin{array}{ccc}
0&A_R&-A_R\\
A_R&B_R&C_R\\
-A_R&C_R&B_R
\end{array}
\right),
\label{form1}
\end{equation}
then $M_{light}$ in Eq.(\ref{form}) is given by
\begin{eqnarray}
A&=&{A_D^2\over A_R}\nonumber\\
B&=&-{B_R-C_R \over {2A_R^2}}A_D^2+{B_D-C_D \over A_R}A_D+{(B_D+C_D)^2\over 
2(B_R+C_R)},\\
C&=&{B_R-C_R\over 2A_R^2}A_D^2-{B_D-C_D\over A_R}A_D+{(B_D+C_D)^2\over 
2(B_R+C_R)}\nonumber
\label{form2}
\end{eqnarray}
This matrix structure is different from that of the quark mass matrix by 
Fritzsch \cite{Fritzsch}, though there is no need for these to coincide.  Then 
there arises a question to what extent this matrix structure Eq.(\ref{form1}) is 
unique under the following assumption: 
\par
(a) $M_{light}$, $M_D$ and $M_R$ have the same structure and that
\par
(b) Their $(1,1)$ components are zeros.
\par
In (a) their structure is not necessarilly identical to Eq.(\ref{form1}). 
%This point will be checked by Super Kamiokande \cite{kajita}.
\par
Running the remaining components of $M_D$ and $M_R$ as free parameters, the 
seesaw mechanism (\ref{light M}) under the conditions (a) and (b) constrains the 
allowed mass matrix in the following four types.
\par
\begin{eqnarray}
(1) \left(
\begin{array}{ccc}
0&A&A\\
A&B&C\\
A&C&C
\end{array}
\right) \quad
(2) 
\left(
\begin{array}{ccc}
0&A&A\\
A&B&B\\
A&B&C
\end{array}
\right) \quad
(3) \left(
\begin{array}{ccc}
0&A&A\\
A&B&C\\
A&C&B
\end{array}
\right)
\quad (4)
\left(
\begin{array}{ccc}
0&A&-A\\
A&B&C\\
-A&C&B
\end{array}
\right) \nonumber
\end{eqnarray}  
(4) is the structure mentioned above.  (3) is transformed to (4) by the 
interchange of $C$ to $-C$ and these are physically equivalent as follows. So 
far we have set $\theta_2$ to be ${\pi \over 4}$. If we leave $\theta_2$ as a 
free parameter and keep the assumption (11), then $M_{light}$ is reduced to
\begin{equation}
\left(
\begin{array}{ccc}
0&c_2\sqrt{m_1m_2}&-s_2\sqrt{m_1m_2}\\
c_2\sqrt{m_1m_2}&(-m_1+m_2)c_2^2+m_3s_2^2&(m_1-m_2+m_3)c_2s_2\\
-s_2\sqrt{m_1m_2}&(m_1-m_2+m_3)c_2s_2&(-m_1+m_2)s_2^2+m_3c_2^2
\end{array}
\right)
\end{equation}
Therefore (3) and (4) are corresponding to $s_2=-{\pi\over 4}$ and  
$s_2={\pi\over 4}$, respectively.  $\theta_2$ has been determined from the 
mixing factor $sin^2 2\theta_2\sim 1$ and they are equivalent.  (1) and (2) are 
also substantially same and are enforced to $m_3=0$. $m_3$ is the heaviest 
neutrino mass and (1) and (2) are rejected.  Thus we obtain the unique structure 
(14) provided that we adopt the assumptions (a) and (b).
\par
Unfortunately our assumptions are not sufficient to realize Eq.(1) 
straightforwardly since Eqs.(1), (12) and (15) can not fix the parameters 
{A,B,C} with subscript D and R. 
%However, these parameters can be made fine tuned to realize Eq.(1). Of course, 
%$f_{i,j}^{(l)}\phi_1=diag(m_e,m_{\mu},m_{\tau})$ also must be satisfied in this 
%case. 
\par
Finally we consider the physical consequences of $M_{light}$ in Eq.(11), 
especially to the neutrinoless double beta decay.
\begin{center}
------------\\
   Fig. 1   \\
------------
\end{center}
The amplitude of this process is proportional to $
<m_{\nu}>$ defined by \cite{Doi}
\begin{equation}
<m_{\nu}>\equiv |\sum _{j=1}^{3}U_{1j}^2m_j|
\end{equation}
Here $U_{ij}$ is in our theory given by Eq.(8) and
\begin{equation}
<m_\nu >=|-c_1^2m_1+s_1^2m_2|
\end{equation}
From the solar neutrino experiment $m_2$ is estimated to be $m_2\sim
O(10^{-3}eV)$, whereas the experimental upper bound of $<m_\nu>$ is of the 
several $eV$ order.  Therefore our estimation is lower than the present upper 
boud by at least $10^{-3}$ times.

\section*{Acknowledgements}

We are grateful to K. Ueda and S. Yu at Ritsumeikan University for useful 
comments.

\vfill\eject

Figure Caption
\par
Fig.1   Feynman diagram of the neutrinoless double beta decay.  For the helicity 
matching of the Majorana neutrino $N_j$ emerges the small factor ${m_j\over E}$.

\vfill\eject

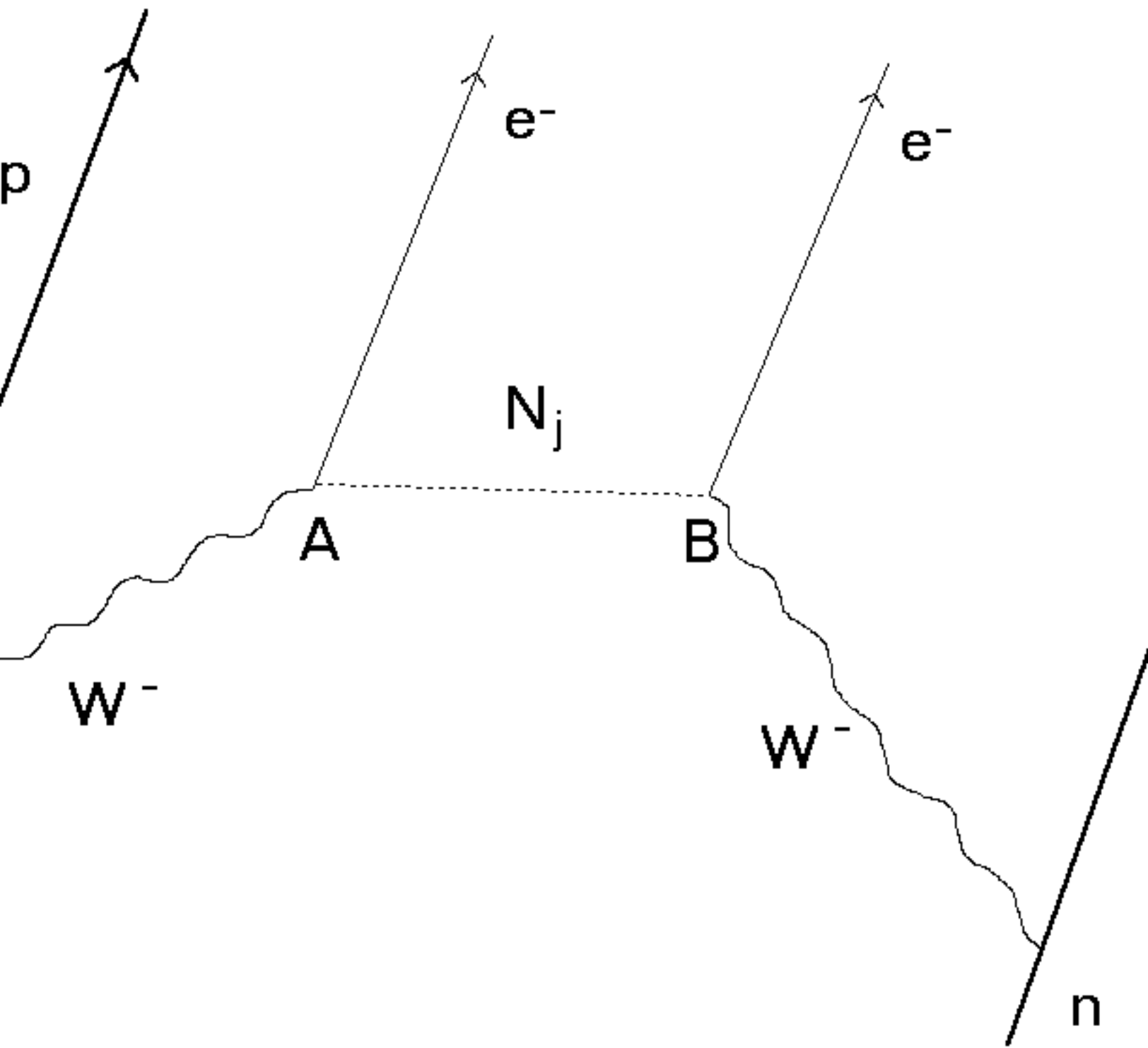
\end{document}